Evolution of superconductivity by oxygen annealing in FeTe$_{0.8}$S$_{0.2}$

Y. Mizuguchi[1,2,3], K. Deguchi[1,2,3], S. Tsuda[1,2], T. Yamaguchi[1,2] and Y. Takano[1,2,3]

1.*National Institute for Materials Science, 1-2-1 Sengen, Tsukuba 305-0047, Japan*

2.*JST, TRIP, 1-2-1 Sengen, Tsukuba 305-0047, Japan*

3.*University of Tsukuba, 1-1-1 Tennodai, Tsukuba 305-0001, Japan*

PACS numbers: 74.70.-b, 74.62.Dh, 74.62.Bf

Abstract

Oxygen annealing dramatically improved the superconducting properties of solid-state-reacted FeTe$_{0.8}$S$_{0.2}$, which showed only a broad onset of superconducting transition just after the synthesis. The zero resistivity appeared and reached 8.5 K by the oxygen annealing at 200 °C. The superconducting volume fraction was also enhanced from 0 to almost 100 %. The lattice constants were compressed by the oxygen annealing, indicating that the evolution of bulk superconductivity in FeTe$_{0.8}$S$_{0.2}$ was correlated to the shrinkage of lattice.

Introduction.

Since the discovery of superconductivity at a transition temperature $T_c$ = 26 K in LaFeAsO$_{1-x}$F$_x$, several types of Fe-based superconductors, essentially composed of 2-dimensional Fe-square lattices, have been discovered [1–6]. Anti-PbO-type FeSe is the simplest Fe-based superconductor [4], and shows comparably high $T_c$ of 37 K under high pressure [7-10]. On the other hand, FeTe is an antiferromagnet with an ordering temperature of 70 K, while it has a structure quite similar to superconducting FeSe. The substitution of S or Se for the Te site suppresses the antiferromagnetic ordering and achieves superconductivity [11-14]. The FeTe$_{1-x}$Se$_x$ superconductors have provided several information useful for understanding the mechanism of Fe-based superconductivity, because the large single crystals could be grown easily. In contrast, the synthesis of the high quality FeTe$_{1-x}$S$_x$ sample is difficult, due to the solubility limit of S for the Te site [14,15].

The superconducting properties of FeTe$_{1-x}$S$_x$ depend on the synthesis method. FeTe$_{0.8}$S$_{0.2}$ synthesized by a melting method showed superconductivity at $T_c^{zero}$ = 7.8 K, but it contained impurity phases and its superconducting volume fraction was only 20 %. The nearly single-phase sample of FeTe$_{0.8}$S$_{0.2}$ was obtained by using the solid-state reaction method. However, the sample showed only filamentary superconductivity and zero resistivity was not observed, due to an insufficiency of S concentration. To elucidate the intrinsic properties of the FeTe$_{1-x}$S$_x$ superconductor, achieving bulk superconductivity in the single-phase sample is strongly required. Recently, we reported moisture-induced superconductivity in FeTe$_{0.8}$S$_{0.2}$ which was synthesized by the solid-state reaction method and showed only filamentary superconductivity just after synthesis [16]. At room temperature, only the sample immersed into the water for several days showed an evolution of bulk superconductivity. Furthermore, the sample immersed into the hot water showed superconductivity with a superconducting volume fraction above 15 % in only 1 day, indicating the speed of evolution of superconductivity was strongly enhanced by heating. In this respect, we investigated the effect of annealing at several temperatures in oxygen gas for as-grown FeTe$_{0.8}$S$_{0.2}$, and found that the FeTe$_{0.8}$S$_{0.2}$ sample annealed at 200 °C shows bulk superconductivity with a sharp superconducting transition.

Experimental methods.
The polycrystalline samples of FeTe$_{0.8}$S$_{0.2}$ were prepared using the solid-state reaction method as described in Refs. 13 and 16. The obtained samples were quickly sealed into the quartz tubes filled with oxygen gas of atmospheric pressure, and annealed at 100, 200, 300 and 400 ºC for 2 hours, respectively. The samples were characterized by the x-ray diffraction using a Cu-Kα radiation. Temperature dependence of resistivity was measured using a four-terminals method from 300 to 2 K. Temperature dependence of magnetic susceptibility was measured using a SQUID magnetometer in a zero-field-cooling (ZFC) mode with an applied magnetic field of 10 Oe. Here we define the sample name as the annealed temperature; for example, $T_a$ = 100 ºC is the FeTe$_{0.8}$S$_{0.2}$ sample annealed at 100 ºC for 2 hours.

Results and discussion.

Figure 1 shows the temperature dependence of magnetic susceptibility normalized at 15 K for oxygen-annealed FeTe$_{0.8}$S$_{0.2}$. For the as-grown sample, no sign of superconductivity was observed. With annealing the sample at 100 ºC, the superconducting transition appeared around 7 K. The sharp superconducting transition at 9.0 K was observed for $T_a$ = 200 ºC, and the superconducting volume fraction reached almost 100 %. Both $T_c$ and superconducting volume fraction decreased for $T_a$ = 300 ºC. Furthermore, superconductivity was not observed for $T_a$ = 400 ºC, indicating that the annealing above 300 ºC was disadvantageous for inducing superconductivity.

Figure 2(a) shows the temperature dependence of resistivity below 15 K for the as-grown sample and $T_a$ = 100, 200 and 300 ºC. The as-grown sample showed only an onset of the superconducting transition around 5 K. By annealing at 100 ºC, the superconducting transition became sharper and the zero resistivity state appeared at 4.3 K. For $T_a$ = 200 ºC, a sharp superconducting transition with $T_c^{onset}$ = 10.7 K and $T_c^{zero}$ = 8.5 K was observed. With annealing at higher temperature of 300 ºC, the $T_c$ was suppressed, corresponding to the result in the magnetic susceptibility measurements. The temperature dependence of resistivity from 300 to 2 K for all samples was shown in Fig. 2(b). For $T_a$ = 100 ºC, the temperature dependence of resistivity showed an increase with decreasing temperature above $T_c$. On the other hand, the temperature dependence of resistivity for $T_a$ = 200 ºC exhibited a broad hump around 60 K and a decrease above $T_c$. The behavior is quite similar to that of the optimally doped FeTe$_{1-x}$Se$_x$, indicating that the oxygen annealing optimized superconductivity for FeTe$_{0.8}$S$_{0.2}$. The annealing at both 300 and 400 ºC increased the resistivity. It would be due to a decomposition of the PbO structure and an increase of impurity phases such as FeTe$_2$.

To investigate the superconducting properties of inside of the annealed samples, the surface of the samples were ground down after the first resistivity measurement. After trimming the sample to less than 1/5 of the pellet, the new terminals were fabricated, and the resistivity was measured again with the same condition. Figure 3 shows the temperature dependence of normalized resistivity for the as-annealed and the trimmed sample with $T_a$ = 100 and 200 ºC. For both samples, no difference of resistivity was observed between the surface and the inside, indicating a homogeneity of the oxygen-annealed sample.

Figure 4(a) shows the temperature dependence of resistivity for $T_a$ = 200 ºC under the magnetic field up to 7 T with an increment of 1 T. Both the upper critical field $\mu_0 H_{c2}$ and irreversible field $\mu_0 H_{irr}$ were plotted in Fig. 4(b) as a function of temperature. Due to the high critical fields, the changes of the onset temperature under magnetic fields were not obvious. Therefore, the $T_c^{onset}$ was defined by a cross point of two lines described in the inset of Fig. 4(a). By extrapolating these data points as described in Fig. 4(b), $\mu_0 H_{c2}(0)$ and $\mu_0 H_{irr}(0)$ were estimated to be ~100 T and ~60 T. By using the WHH theory [17], $\mu_0 H_{c2}(0)$ was calculated to be ~ 70 T. These values are almost the same as the previous report on the $FeTe_{0.8}S_{0.2}$ synthesized by the melting method. Oxygen-annealed $FeTe_{0.8}S_{0.2}$ is one of the candidate materials for application, due to its high $\mu_0 H_{c2}$ and the homogeneity of the annealed sample.

To clarify the origin of the evolution of bulk superconductivity, the x-ray diffraction measurements were performed for the oxygen-annealed samples. The typical diffraction peaks of (101) and (002) are shown in Fig. 5(a). The peaks shifted to higher angles for $T_a$ = 100, 200 and 300 ºC compared to the as-grown sample, indicating that the oxygen annealing compressed the lattice. With annealing at 400 ºC, the peaks shifted to lower angles and the peaks of the impurity phases grew up. The calculated lattice constants $a$ and $c$ were plotted in Fig. 5(b) and (c), respectively, as a function of annealing temperature. For comparison, we also plotted the lattice constants of $FeTe_{0.8}S_{0.2}$ synthesized by the melting method, which showed superconductivity with a superconducting volume fraction of ~20 %. The lattice constants were clearly compressed for $T_a$ = 100, 200 and 300 ºC compared to the as-grown sample. In particular, the $a$ axis of the oxygen-annealed samples were smaller than that of $FeTe_{0.8}S_{0.2}$ synthesized by the melting method. These results suggest that bulk superconductivity of $FeTe_{1-x}S_x$ can be induced when the lattice is optimally compressed.

To investigate the effects of the annealing in several atmospheres other than oxygen, the as-grown samples were sealed into an evacuated tube and tubes filled with nitrogen and argon gas, respectively, and annealed at 200 ºC for 2 hours. Figure 6(a) shows the temperature dependence of magnetic susceptibility normalized at 15 K for those samples. Superconductivity was not observed for the samples annealed in vacuum, nitrogen and argon. We also investigated the oxygen-annealing effect for polycrystalline FeTe, which shows antiferromagnetic/structural transition around 70 K. Figure 6(b) shows the temperature dependence of magnetic susceptibility for FeTe and FeTe$_{0.8}$S$_{0.2}$ both annealed in oxygen at 200 ºC for 2 hours. There was no sign of superconductivity for oxygen-annealed FeTe. Although oxygen-incorporation-induced filamentary superconductivity in a FeTe film was recently reported [18], superconductivity was not observed for the bulk FeTe sample annealed in oxygen. In this respect, the oxygen annealing is effective to induce bulk superconductivity only for FeTe$_{1-x}$S$_x$.

Conclusion.

In conclusion, the oxygen annealing dramatically improved superconducting properties for solid-state-reacted FeTe$_{0.8}$S$_{0.2}$, which showed only filamentary superconductivity just after the synthesis. The zero resistivity appeared and reached 8.5 K by the oxygen annealing at 200 °C. The superconducting volume fraction was also enhanced from 0 to almost 100 %. High upper critical field of 70 T was estimated for oxygen-annealed FeTe$_{0.8}$S$_{0.2}$. Because of the high critical field and the homogeneity of the annealed sample, oxygen-annealed FeTe$_{0.8}$S$_{0.2}$ is one of the candidate materials for application. The lattice constants were compressed by the oxygen annealing, indicating that the evolution of bulk superconductivity was correlated to the shrinkage of the lattice. To elucidate the accurate oxygen site introduced by the oxygen annealing, the detailed structural analysis and/or the measurements sensitive to the local structure should be performed.

Acknowledgement.

This work was partly supported by Grant-in-Aid for Scientific Research (KAKENHI).

Figure captions

Fig. 1. (Colour on-line) Temperature dependence of magnetic susceptibility for as-grown FeTe$_{0.8}$S$_{0.2}$ and $T_a$ = 100, 200, 300 and 400 ºC. The susceptibility was normalized at 15 K to compare all data.

Fig. 2. (Colour on-line) (a)Temperature dependence of resistivity below 15 K for the as-grown sample and $T_a$ = 100, 200 and 300 ºC. (b)Temperature dependence of resistivity from 300 to 2 K for the as-grown sample and $T_a$ = 100, 200, 300 and 400 ºC.

Fig. 3. (Colour on-line) Temperature dependence of normalized resistivity measured with four terminals attached to the surface of the as-annealed pellet and the trimmed surface (inside of the pellet) for $T_a$ = 100 and 200 ºC.

Fig. 4. (Colour on-line) (a)Temperature dependence of resistivity for $T_a$ = 200 ºC under magnetic fields up to 7 T with an increment of 1 T. The inset shows an enlargement of the superconducting transition at 0 T and definition of $T_c^{onset}$ and $T_c^{zero}$. (b)Temperature dependence of $\mu_0 H_{c2}$ and $\mu_0 H_{irr}$ for $T_a$ = 200 ºC. The dash lines are the linear extrapolations of the data points.

Fig. 5. (Colour on-line) (a) Typical diffraction peaks of (101) and (002) for the as-grown sample and $T_a$ = 100, 200, 300 and 400 ºC. (b), (c)Annealing temperature dependence of lattice constants $a$ and $c$. Blue open circles are the lattice constants of FeTe$_{0.8}$S$_{0.2}$ synthesized by the melting method.[14]

Fig. 6. (Colour on-line) (a) Temperature dependence of normalized magnetic susceptibility for the FeTe$_{0.8}$S$_{0.2}$ samples annealed in vacuum, nitrogen and argon. (b) Temperature dependence of normalized magnetic susceptibility for FeTe$_{0.8}$S$_{0.2}$ and FeTe both annealed in oxygen at 200 ºC for 2 hours.

Fig. 1

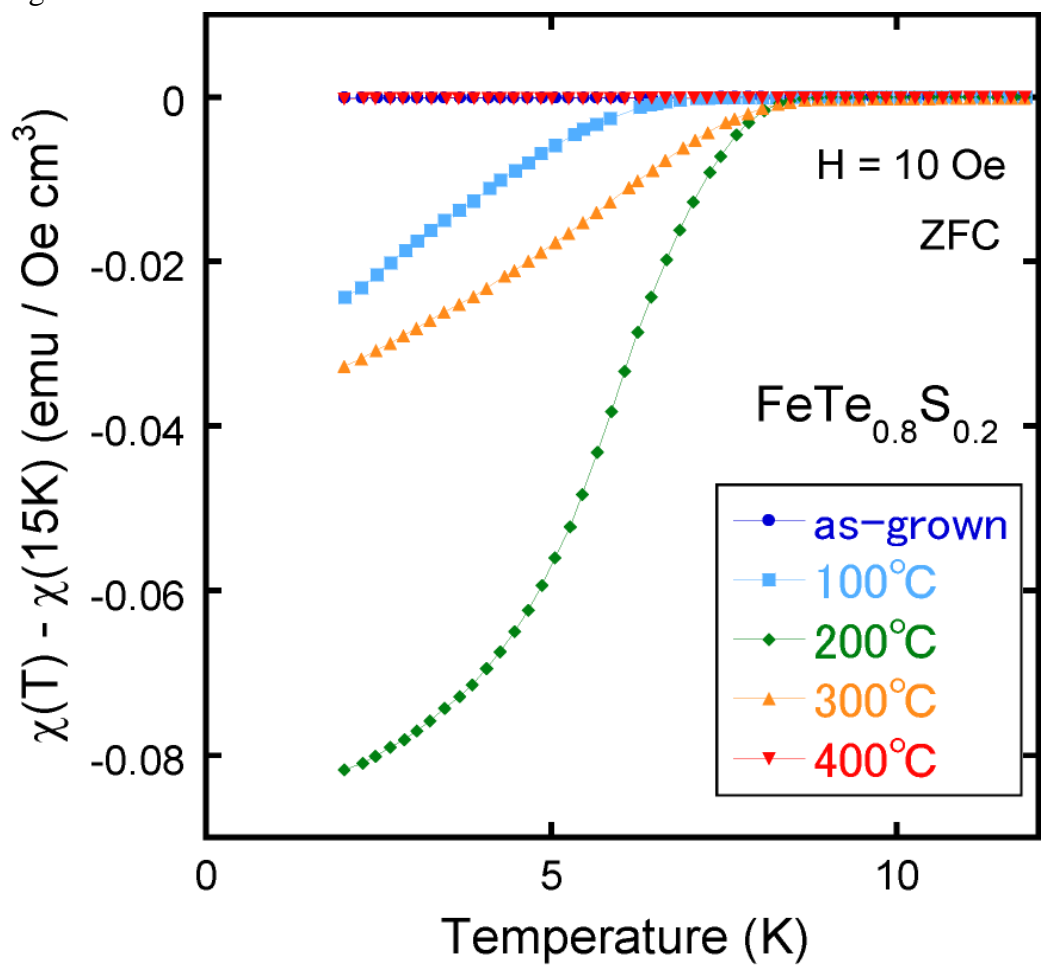

Fig. 2

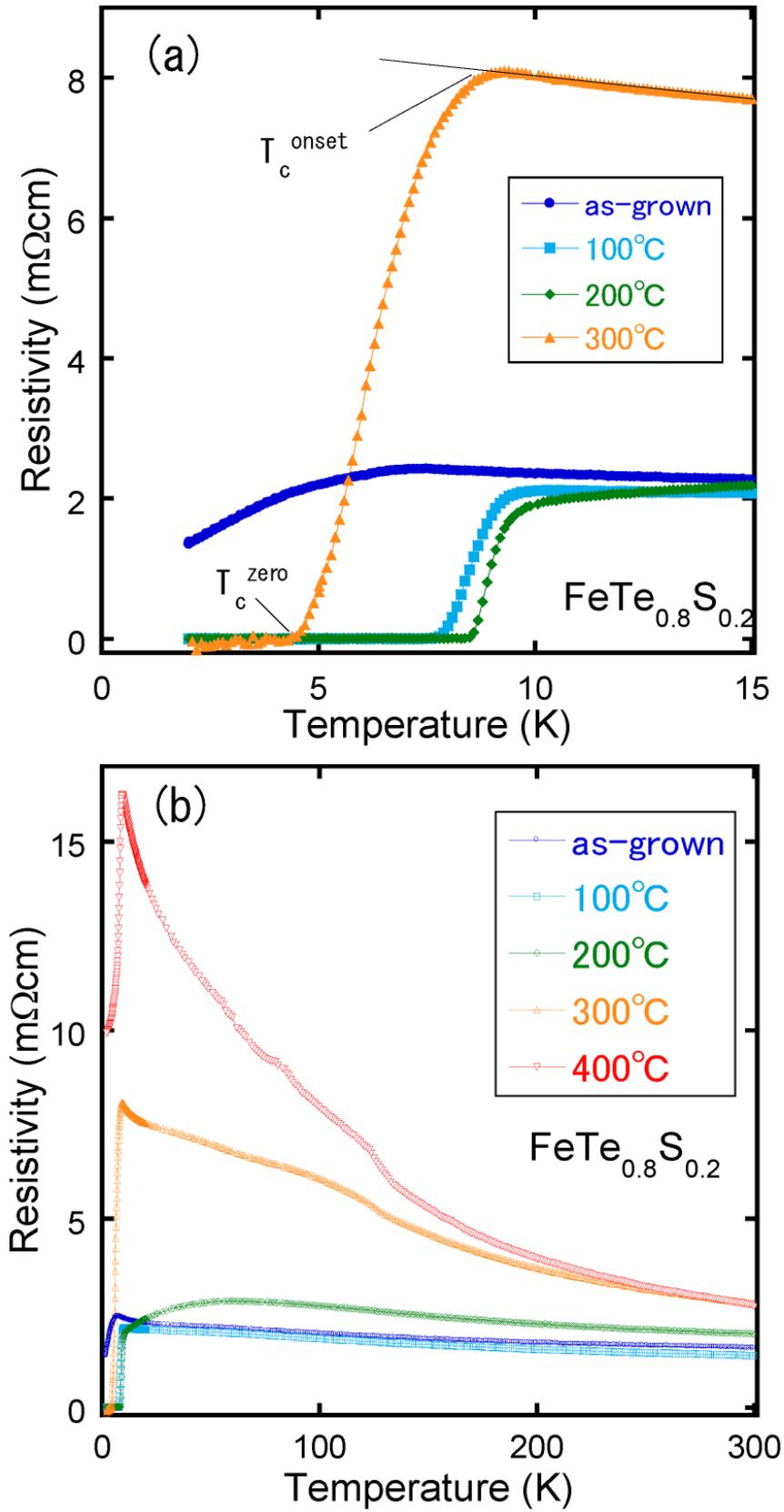

Fig. 3

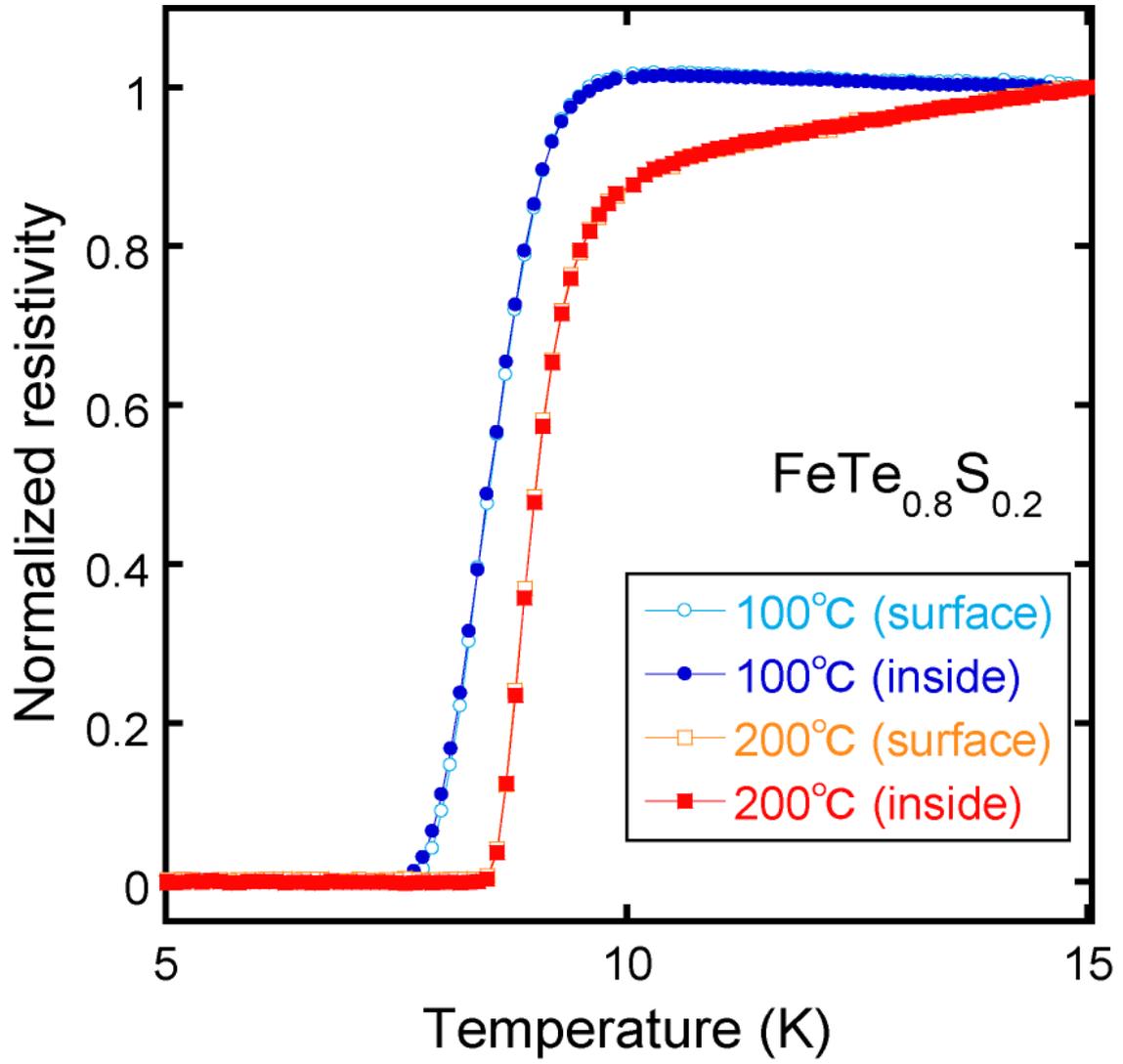

Fig. 4

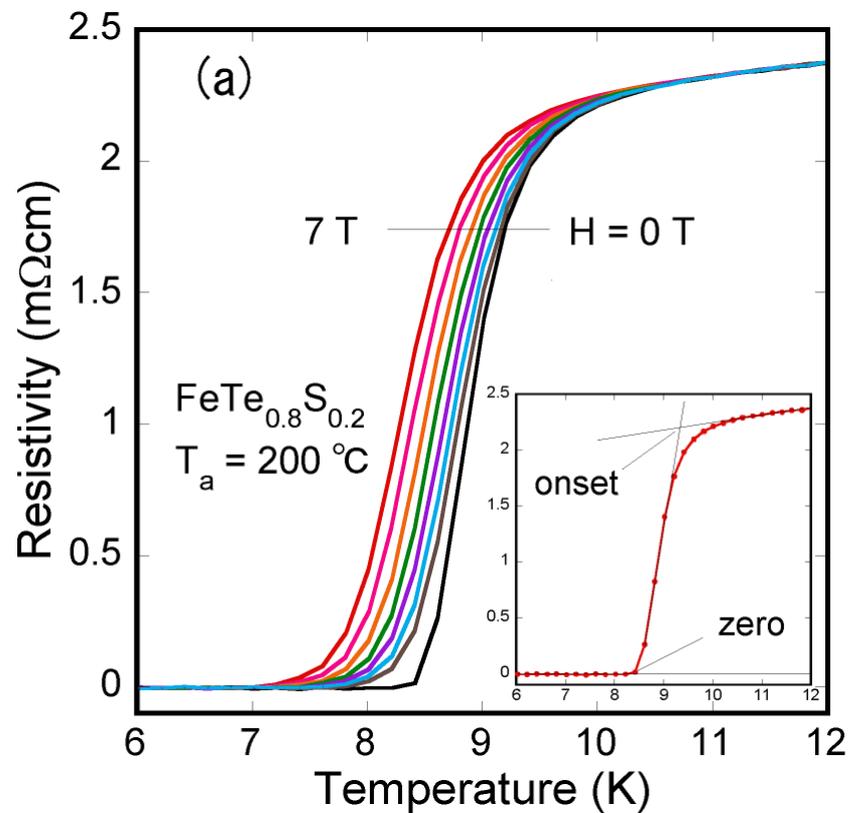

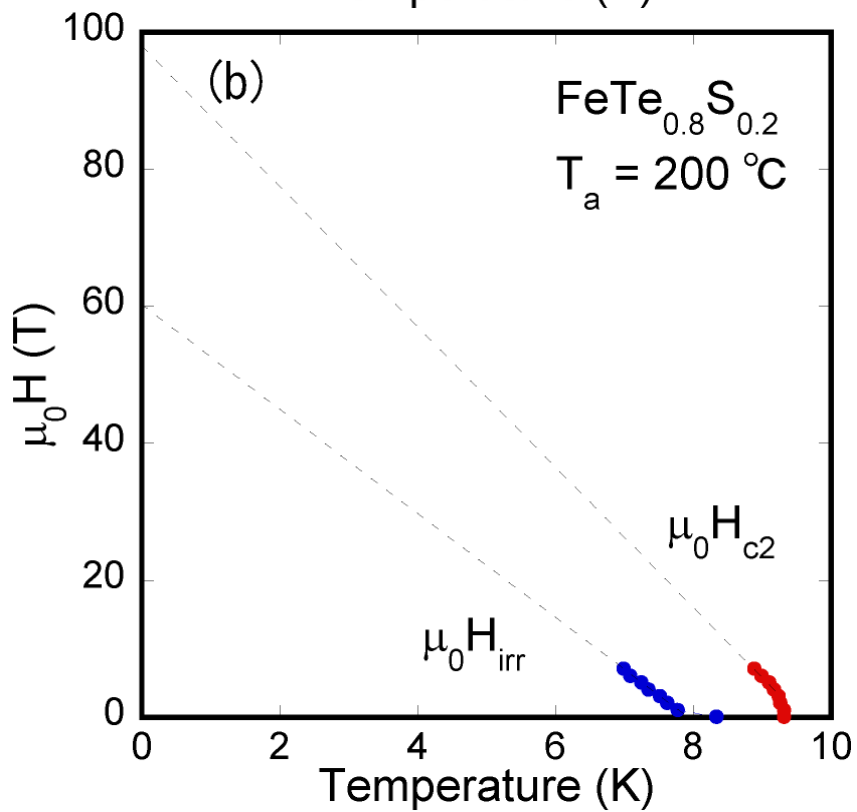

Fig. 5

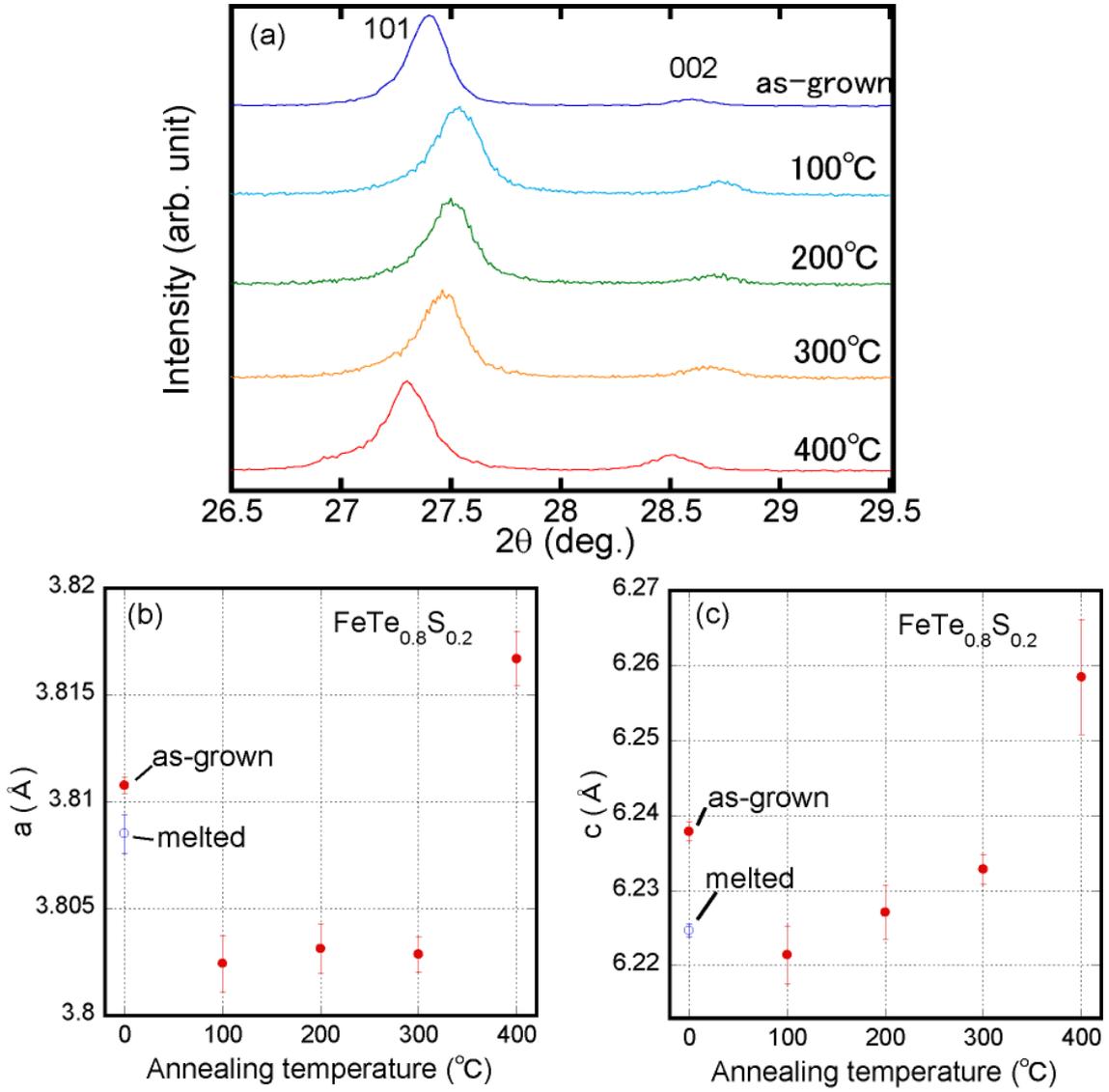

Fig. 6

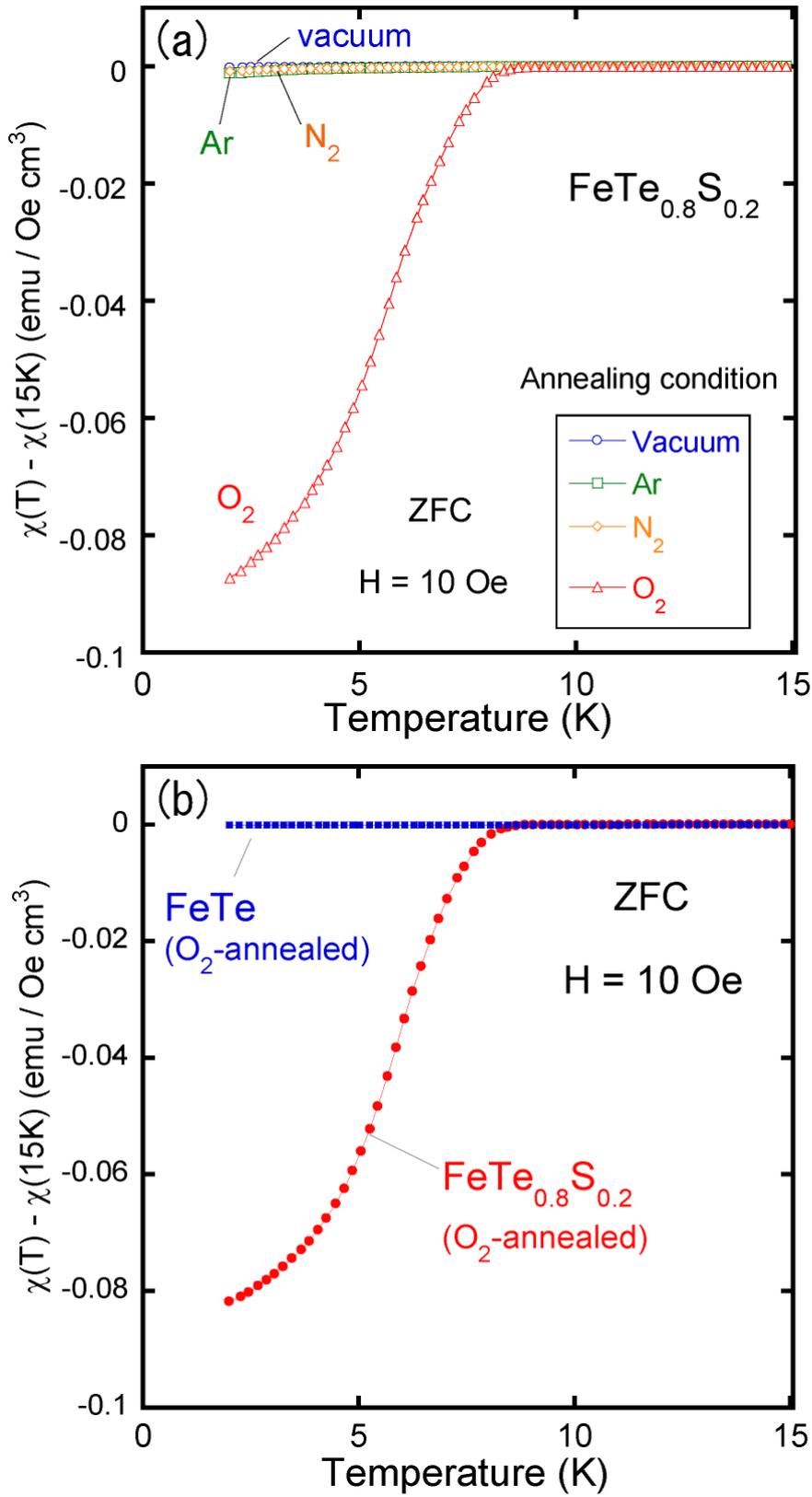